\documentclass[12pt]{article}
\usepackage{graphicx}
\usepackage{color}
\usepackage{cite}

\def \beq{\begin{equation}}
\def \eeq{\end{equation}}
\def\eqref#1{(\ref{#1})}
\def\bea{\begin{eqnarray}}
\def\eea{\end{eqnarray}}
\def\jpsi{J\kern-0.15em/\kern-0.15em\psi\kern0.15em}

\def\URLtilde{\lower0.2em\hbox{$\tilde{\phantom{a}}$}}
\def\mycomm#1{\hfill\break\strut\kern-3em{\color{red}\tt ====> #1
\color{black}}\hfill\break}

\newcount\timecount
\newcount\hours \newcount\minutes  \newcount\temp \newcount\pmhours
\hours = \time
\divide\hours by 60
\temp = \hours
\multiply\temp by 60
\minutes = \time
\advance\minutes by -\temp
\def\hour{\the\hours}
\def\minute{\ifnum\minutes<10 0\the\minutes
\else\the\minutes\fi}
\def\clock{
\ifnum\hours=0 12:\minute\ AM
\else\ifnum\hours<12 \hour:\minute\ AM
\else\ifnum\hours=12 12:\minute\ PM
\else\ifnum\hours>12
\pmhours=\hours
\advance\pmhours by -12
\the\pmhours:\minute\ PM
\fi
\fi
\fi
\fi
}

\def\monthname{\relax\ifcase\month 0/\or January\or February\or
March\or April\or May\or June\or July\or August\or September\or
October\or November\or December\else\number\month/\fi}

\def\bold#1{\setbox0=\hbox{$#1$}     \kern-.025em\copy0\kern-\wd0
\kern.05em\copy0\kern-\wd0
\kern-.025em\raise.0433em\box0 }

\begin{document}
\setcounter{footnote}{1}
\rightline{EFI 14-39}
\rightline{TAUP 2989/14}
\rightline{arXiv:1410.7729}
\vskip1.5cm

\centerline{\large \bf \boldmath $X(3872)$, $X_b$, and the $\chi_{b1}(3P)$
state \unboldmath}
\bigskip

\centerline{Marek Karliner$^a$\footnote{{\tt marek@proton.tau.ac.il}}
 and Jonathan L. Rosner$^b$\footnote{{\tt rosner@hep.uchicago.edu}}}
\medskip

\centerline{$^a$ {\it School of Physics and Astronomy}}
\centerline{\it Raymond and Beverly Sackler Faculty of Exact Sciences}
\centerline{\it Tel Aviv University, Tel Aviv 69978, Israel}
\medskip

\centerline{$^b$ {\it Enrico Fermi Institute and Department of Physics}}
\centerline{\it University of Chicago, 5620 S. Ellis Avenue, Chicago, IL
60637, USA}
\bigskip
\strut

\begin{center}
ABSTRACT
\end{center}
\begin{quote}
We discuss the possible production and discovery channels in $e^+e^-$ and
$pp$ machines of the $X_b$, the bottomonium counterpart of $X(3872)$ and the
putative isoscalar analogue of the charged bottomonium-like states $Z_b$
discovered by Belle.  We suggest that the $X_b$ may be close in mass to the
bottomonium state $\chi_{b1}(3P)$, mixing with it and sharing its decay
channels, just as $X(3872)$ is likely a mixture of a $\bar D D^*$ molecule
and $\chi_{c1}(2P)$.  Consequently, the experiments which reported observing
$\chi_{b1}(3P)$ might have actually discovered the $X_b$, or a mixture of the
two states.
\end{quote}
\smallskip

\leftline{PACS codes: 14.20.Lq, 14.20.Mr, 12.40.Yx}
\bigskip


\section{Introduction \label{sec:intro}}

The search for ``exotic'' mesons made of more than a quark and an antiquark,
and for exotic baryons made of more than three quarks, is almost as old as the
quark model itself \cite{Gell-Mann:1964,Zweig:1964}.  (For an early suggestion
of exotic mesons in baryon-antibaryon channels see Ref.\ \cite{Rosner:1968}.)
However, for many years no such states were conclusively observed \cite{PDG}.
The discovery of a neutral state $X(3872)$ \cite{X3872} in 2003, where 3872
stands for the mass in MeV/$c^2$, suggested the possibility of richer
structures, such as $c \bar c q \bar q$, where $c$ is a charmed quark and $q
= u$ or $d$.  This particle is now most plausibly understood as a mixture of
a P-wave charmonium $\chi_{c1}(2P)$ level of spin 1 and an S-wave molecule of
$D^0 \bar D^{*0} + {\rm c.c.}$ \cite{Xmix} whose binding energy is so close to
zero that its sign is not yet known \cite{PDG}.  Evidence for a charged
counterpart of this particle at a mass of about 3900 MeV came several years
later \cite{Ablikim:2013zna,Liu:2013dau}.

Molecular states of charmonium had been proposed a number of years ago
\cite{Voloshin:1976ap,DGG,deusons}.  The fact that the $X(3872)$ can decay
both to $J/\psi \rho^0$ and $J/\psi \omega$ is accounted for by the
substantial isospin splitting between $D^0 \bar D^{*0} + {\rm c.c.}$ and
$D^+ D^{*-} + {\rm c.c.}$, so that the tetraquark configuration of
$X(3872)$ is mainly $c \bar c u \bar u$, permitting equal coupling to
$\rho^0$ and $\omega$ \cite{Tornqvist:2004qy}.

It was proposed in 2008 \cite{Karliner:2008rc} that similar behavior in the
bottomonium system could lead to resonant effects near $B \bar B^*+{\rm c.c.}$
thresholds.  (See also Refs.\ \cite{Karliner:2011yb,Karliner:2012pc}.)
In 2011 a charged candidate for such a state, $Z_b(10610)$,
along with a candidate for a $B^* \bar B^*$ state $Z_b(10650)$, was observed
by the Belle Collaboration \cite{Belle:2011aa}.

In the present article we propose ways to look for the $X_b$, which plays the
dual role of the bottomonium analogue of the $X(3872)$ and the isoscalar partner
of the $Z_b(10610)$.  In Section II we discuss some expected general features 
of $X_b$, including estimates of its mass if it is a bound state of $B \bar B^*
+{\rm c.c.}$, and a discussion of its likely mixing with a nearby bottomonium
state, the recently observed P-wave $b \bar b$ excitation $\chi_{b1}(3P)$ 
\cite{ATLASchi3P,D0chi3P,LHCbchi3Pa,LHCbchi3Pb}.  Properties of the
$\chi_{b1}(3P)$, especially its radiative decays, are treated in Sec.\ III.  A 
corresponding discussion for the $\chi_{c1}(2P)$, expected to mix with the
$X(3872)$, is given in Sec.\ IV.  The isoscalar nature of the $X_b$ is the
subject of Sec.\ V.  Consequences for observing the $X_b$
are noted in Sec.\ VI, while Sec.\ VII concludes.
\hfill\break

\section{Expected general features of $X_b$}

An accurate estimate of the mass of $X(3872)$ \cite{deusons} was made long ago
based on a calculation of the binding of $D^0$ and $\bar D^{*0}$ due to pion
exchange and other forces.  This estimate was extended to bottom mesons,
leading to a predicted mass $M(X_b)=10562$ MeV \cite{deusons,Tornqvist:2004qy}
for a $B \bar B^* + {\rm c.c.}$ state of $J^{PC} = 1^{++}$.  This mass is quite
close to that observed for the P-wave $b \bar b$ excitation $\chi_{b1}(3P)$,
which ranges from about 10510 to 10550 MeV/$c^2$.  One prediction
\cite{Kwong:1988ae} estimates $M(\chi_{b1}(3P)) = 10516$ MeV/$c^2$ but the
spread in models is modest as the interquark potential is well known in this
regime \cite{Quigg:1981bj}.

An independent estimate of $M(X_b)$, to be mentioned in more detail below,
was based on the expected binding energy of a $B$ and $\bar B^*$, yielding
10585 MeV/$c^2$ \cite{Karliner:2013dqa}.  This is 23 MeV above Tornqvist's
value, but still fairly close to the expected mass of $\chi_{b1}(3P)$.  Thus,
in either case, there should be two nearby states with $I=0,\,J^{PC} = 1^{++}$,
sharing common decay modes to some extent.  Another estimate, appearing in an
unpublished version of Ref.\ \cite{AlFiky:2005jd}, is $M(X_b) = 10604$
MeV/$c^2$, just below $B \bar B^*$ threshold.

The radiative decay widths of a $B \bar B^* + {\rm c.c.}$ bound state to
$\Upsilon(1S,2S,3S)$ have been estimated to be quite small, of order
(1.5,1.8,0.4) keV or less \cite{Li:2014uia}.  Unfortunately they depend
sensitively on unknown parameters; a set favored by the authors predicts
(0.7,0.5,0.2) keV for these values.  Note that the decay width to the 3S
state is smaller than those to 1S or 2S.  As we shall see, only a small
admixture of $\chi_{b1}(3P)$ in the $B \bar B^* + {\rm c.c.}$ wave function
can alter this pattern drastically.

\section{Properties of the $\chi_{b1}(3P)$ bottomonium state}

\subsection{Mass}

The reported mass values of the $\chi_{b1}(3P)$
are summarized in Table \ref{tab:m3P}.  These are compatible
with the predicted value of 10516 MeV/$c^2$ in Ref.\ \cite{Kwong:1988ae}.
The differences between the first two values and the last two, if not due
to limited statistics, may stem from different admixtures of $\chi_{b2}(3P)$
in central \cite{ATLASchi3P,D0chi3P} and forward \cite{LHCbchi3Pa,LHCbchi3Pb}
production.


\begin{table}
\caption{Values of $M(\chi_{b1}(3P))$ observed in various experiments.
\label{tab:m3P}}
\begin{center}
\begin{tabular}{c c c} \hline \hline
Collaboration & Reference & Value (MeV/$c^2$) \\ \hline
  ATLAS  & \cite{ATLASchi3P} & $10530 \pm 5 \pm 9$ \\
    D0   & \cite{D0chi3P}    & $10551 \pm 14 \pm 17$ \\
LHCb (a) & \cite{LHCbchi3Pa} & $10511.3 \pm 1.7 \pm 2.5$ \\
LHCb (b) & \cite{LHCbchi3Pb} & $10515.7^{+2.2+1.5}_{-3.9-2.1}$ \\ \hline \hline
\end{tabular}
\end{center}
\leftline{(a) Using non-converted photons. (b) Using converted photons.}
\end{table}

\subsection{Radiative decays}

A key feature of the $\chi_{b1}(3P)$ is the expected dominance of
$\Gamma(\chi_{b1}(3P) \to \gamma \Upsilon(3S)$) over
$\Gamma(\chi_{b1}(3P) \to \gamma \Upsilon(2S)$) or
$\Gamma(\chi_{b1}(3P) \to \gamma \Upsilon(1S)$, as a result of a much larger
electric dipole amplitude.  The rate for an E1 transition from a $^3P_1$ state
with radial quantum number $n$ to a $^3S_1$ state with radial quantum $n'$ in a
quarkonium $Q\bar Q$ system \cite{Kwong:1988ae,Eichten:1980,Buchmuller:1981,%
Moxhay:1983, Sterling:1978} is 
\beq \label{eqn:rate}
\Gamma(n^3P_1) \to \gamma n'^3S_1) = \frac{4}{9}e_Q^2 \alpha E_\gamma^3
|\langle n'|r|n \rangle|^2~,
\eeq
where $e_Q$ is the charge of $Q$ (2/3 for $c$, --1/3 for $b$), $\alpha =
1/137.036$, and $E_\gamma$ is the photon energy.  Dipole matrix elements
predicted in Ref.\ \cite{Kwong:1988ae}, photon energies, and partial decay
widths are summarized in Table \ref{tab:3Pbb}.
\begin{table}
\caption{Dipole matrix elements, photon energies, and partial decay widths
for the transitions $\chi_{b1}(3P) \to \gamma \Upsilon(n'S)$.
\label{tab:3Pbb}}
\begin{center}
\begin{tabular}{c c c c} \hline \hline
$n'$ & $\langle n'|r|3 \rangle$ & $E_\gamma$ & $\Gamma(\chi_{b1}(3P)) \to$ \\
     & (GeV)$^{-1}$ & (MeV) & $\gamma \Upsilon(n'S))$ (keV) \\ \hline
  1  & 0.101 & 1003 &  3.69 \\
  2  & 0.298 &  481 &  3.56 \\
  3  & 2.627 &  159 &  10.1 \\ \hline \hline
\end{tabular}
\end{center}
\end{table}
The dominance of the transition to the 3S level is a key signature that one
is dealing with a state with at least a substantial admixture of
$\chi_{b1}(3P)$.  Indeed, LHCb has observed the transition $\chi_b(3P) \to
\gamma \Upsilon(3S)$ \cite{LHCbchi3Pa}, but ratios of branching fractions
of $\chi_b(3P) \to \gamma \Upsilon(1S,2S,3S)$ are not quoted.

\section{Properties of the $\chi_{c1}(2P)$}

The hierarchy of electric dipole matrix elements, discussed in the previous
Section for bottomonium transitions, also applies to charmonium transitions,
and is likely the reason that the branching ratio of $X(3872)$ to $\gamma
\psi(2S)$ is larger than that for $X(3872) \to \gamma J/\psi$.

We assume that $X(3872)$ is a mixture of the charmonium state $\chi_{c1}(2P)$
and a $D^0 \bar D^{*0} + {\rm c.c.}$ molecule \cite{Xmix,Dong:2009uf}.  This
assumption is supported by a measurement by the LHCb Collaboration
\cite{Aaij:2014ala} of the ratio
\beq \label{eqn:R}
R_{\psi \gamma} \equiv \frac{{\cal B}(X(3872) \to \psi(2S) \gamma)}
{{\cal B}(X(3872) \to J/\psi \gamma)} = 2.46 \pm 0.64 \pm 0.29~.
\eeq
These authors quote numerous theoretical predictions for $R_{\psi \gamma}$:
(3--4)$\times 10^{-3}$ for a $D \bar D^*$ molecule (see, e.g.,
\cite{Dong:2009uf}), 1.2--15 for a pure charmonium state, and 1.5--5 for a
molecule-charmonium admixture.  (Ref.\ \cite{Chen:2013upa} addresses the
question of how to determine the relative fractions of $c \bar c$ and
$D \bar D^*$ molecule in $X(3872)$ using the line shape in the decay
$X(3872) \to (D^0 \bar D^{*0} + {\rm c.c.})$.)  The large
variation for charmonium is due mainly to uncertainty in the size of the
electric dipole matrix element $\langle 1S|r|2P \rangle$, which is sensitive to
cancellations between contributions from positive and negative values of the
$2P$ wave function.  Such a matrix element would vanish completely in a
harmonic oscillator potential.  The pattern of such suppressions has been
discussed in \cite{Grant:1992fi}.

We estimate the decay rates of a pure charmonium state at 3872 MeV.  We found
$\langle 1S|r|2P \rangle = 0.240$ GeV$^{-1}$ and $\langle 2S|r|2P \rangle =
1.911$ GeV$^{-1}$ for the bottomonium system in Ref.\ \cite{Kwong:1988ae}.
Using the similarity of the charmonium and bottomonium interaction to a
logarithmic potential \cite{Quigg:1977dd}, one can then obtain $\langle 1S|r|2P
\rangle$ and $\langle 2S|r|2P \rangle$ for charmonium by a simple rescaling
\cite{Quigg:1979vr}.  For a $Q \bar Q$ bound state of quarks with mass $m_Q$ in
a potential $V(r) = \lambda r^a$, lengths scale as $\ell \sim m_Q^{-1/(2+a)}$,
and hence as $\ell^{-1/2}$ in a logarithmic potential.  We thus obtain
$\langle 1S|r|2P \rangle = 0.416$ GeV$^{-1}$ and $\langle 2S|r|2P \rangle =
3.315$ GeV$^{-1}$ for charmonium, using the quark masses $m_c = 1663$ MeV
and $m_b = 5004$ MeV from fits to spectra of mesons with one $c$ or $b$ quark
\cite{Karliner:2014gca}.

\begin{table}
\caption{Comparison of electric dipole matrix elements and decay rates for
$\chi_{c1}(2P) \to (J/\psi,\psi(2S)) \gamma$ transitions in various
treatments.
\label{tab:2PtoS}}
\begin{center}
\begin{tabular}{c c c c c c} \hline \hline
Reference & $\langle 1S|r|2P \rangle$ & $\Gamma(\chi_{c1}(2P) \to$ &
 $\langle 2S|r|2P \rangle$ & $\Gamma(\chi_{c1}(2P) \to$ &
 $R(\chi_{c1}(2P))$ \\
 &(GeV)$^{-1}$&$J/\psi \gamma)$ (keV) &(GeV)$^{-1}$ & $\psi(2S) \gamma)$ (keV)
 & \\ \hline
This work & 0.416 & 84.7 & 3.315 & 94.1 & 1.11 \\
\cite{Barnes:2005pb} & 0.348 & 59.2 & 3.196 & 87.5 & 1.48 \\
\cite{Barnes:2003vb} & 0.150 & 11.0 & 2.723 & 63.9 & 5.81 \\
\cite{Li:2009zu} & 0.412 & 83 & 3.468 & 103 & 1.24 \\
\cite{Li:2009zu} & (a) & 45 & (a) & 60 & 1.33 \\ 
\cite{Wang:2010ej} & 0.260 & 33 & 4.128 & 146 & 4.42 \\
\cite{Ferretti:2014xqa} & 0.150 & 11.0 & 2.859 & 70 & 6.36 \\
\hline \hline
\end{tabular}
\end{center}
\leftline{(a) Relativistic calculation}
\end{table}

Using Eq.\ (\ref{eqn:rate}) for electric dipole transitions from a $n^3P_1$ to
a $n'^3S_1$ state, and photon energies of (697,181) MeV for the transitions
from 3872 MeV to (3097,3686) MeV, we then predict
\beq
\Gamma(\chi_{c1}(2P) \to J/\psi \gamma) = 84.7~{\rm keV}~;~~
\Gamma(\chi_{c1}(2P) \to \psi(2S) \gamma) = 94.1~{\rm keV}~;
R(\chi_{c1}(2P)) = 1.11~,
\eeq
where $R(\chi_{c1}(2P))$ is the ratio in Eq.\ (\ref{eqn:R}) for a pure
$\chi_{c1}(2P)$.  These values are compared with some others
\cite{Barnes: 2005pb,Barnes:2003vb,Li:2009zu,Wang:2010ej,Ferretti:2014xqa}
in Table \ref{tab:2PtoS}. (Several
other calculations \cite{Lahde:2002wj,Badalian:2012jz,Mehen:2011ds,%
DeFazio:2008xq}, based on varying assumptions beyond the use of Eq.\ (1),
give values of $R(\chi_{c1}(2P))$ mostly within the same range.)  Where
electric dipole matrix elements are not quoted, we have extracted them from
these references and rescaled predicted widths to photon energies appropriate
for $X(3872)$.  One sees considerably more variation in $\langle 1S|r|2P
\rangle$ than in $\langle 2S|r|2P \rangle$.  One should expect, similarly,
to be less confident about estimates of $\langle 1S|r|3P \rangle$ and
$\langle 2S|r|3P \rangle$ than of $\langle 3S|r|3P \rangle$ in bottomonium.

\section{Isoscalar nature of the $X_b$}

Recently CMS and ATLAS have searched for the decay $X_b \to \Upsilon(1S)\pi^+
\pi^-$ \cite{Chatrchyan:2013mea,Aad:2014ama}. The search in this particular
channel was motivated by the seemingly analogous decay $X(3872) \to \jpsi
\pi^+ \pi^-$.  For this particular decay channel the analogy is misguided
\cite{MKtoCMS,Guo:2014sca}.  The null result of the search described in this
paper does not tell us if the $X_b$ exists, because for an isoscalar 
with $J^{PC} = 1^{++}$ such a decay is forbidden by $G$-parity conservation.

So how come $X(3872)$ which also has $J^{PC}=1^{++}$ {\em does} decay into
$\jpsi \pi^+ \pi^-$ ?  In the charm sector isospin is badly broken between
$D^+$ and $D^0$. $D^+$ is 4.76 MeV heavier than $D^0$, while $D^{*+}$ is 3.30
MeV heavier than $D^{*0}$.  Since $X(3872)$ is right at the $\bar{D} D^*$
threshold, its decays break isospin, so $X(3872) \to \jpsi \pi^+ \pi^-$ is
allowed. In fact, $X(3872)$ decays roughly equally to $\jpsi \rho$ and to
$\jpsi \omega$,  which of course cannot happen if the decays conserve isospin.

On the other hand, in the bottom sector the $B^0- B^+$ mass difference is tiny,
0.32 MeV, so isospin is very well conserved in the decays of $(\bar B B^*)$
``molecules" $X_b$, and the decay $X_b \to \Upsilon \pi^+ \pi^-$ is forbidden.

A simple way to quantify the difference between isospin violation in decays of
$X(3872)$ and $X_b$ is to compare the binding energy with the isospin splitting
in the two-meson sector.  In order for $X(3872)$ to be pure $I=0$, it would
have to be an equal admixture of $\bar{D^0} D^{0*}$ and $D^- D^{+*}$. But the
masses of these two components are very different:

\bea
\label{eq:DDiso}
M(\bar{D^0} D^{0*}) &=& (3871.80 \pm 0.12) \ {\rm MeV},\nonumber\\
vs. \phantom{aaaaaaaaaaaa} &&\\
M(D^- D^{+*}) &=&  (3879.87 \pm 0.12) \ {\rm MeV}.\nonumber
\eea
$X(3872)$, with a quoted mass \cite{PDG} of $(3871.69 \pm 0.17)$ MeV,
is essentially at $\bar{D^0} D^{0*}$ threshold, i.e., the binding
energy is less than 1/2 MeV. The would-be $D^- D^{+*}$ component is well above
threshold and does not contribute. The $\bar{D^0} D^{0*}$ is a combination
of $I=0$ and $I=1$ and therefore $X(3872)$ decays into both 
$\jpsi \omega$ and $\jpsi \rho$ with roughly equal branching fractions.

In Ref.~\cite{Karliner:2013dqa} the $X_b$ binding energy was estimated with the
help of the existing data: $Z_b(10610)$, $Z_c(3900)$ in the $I=1$ channel and
$X(3872)$ in the $I=0$ channel.  Since the kinetic energy is inversely
proportional to mass, the heavier the heavy quark, the deeper the binding.  The
upshot is that even in the most extreme case of infinitely heavy $b$-quark
analogue the binding energy is 35 MeV.  For the real-world 5 GeV $b$-quark, the
binding energy was found to be significantly smaller, about 20 MeV.

With $X_b$ about 20 MeV below $\bar{B} B^*$ threshold the situation in the
bottomonium sector is very different from the charmonium sector:
\bea
\label{eq:BBiso}
M(\bar{B^0} B^{0*}) &=& (10604.8 \pm 0.4) \ {\rm MeV}, \nonumber\\
vs. \phantom{aaaaaaaaaaaa} &&\\
M(B^- B^{+*})    &=& (10604.5 \pm 0.4) \ {\rm MeV}.\nonumber
\eea
In this case the isospin splitting is very small compared with the binding
energy: $(0.3 \pm 0.4)$ MeV vs. at least 20 MeV, i.e., at most 1.5\%. Therefore
$X_b$ will be an almost pure isoscalar.  The estimate of Ref.\
\cite{Karliner:2013dqa} predicts its mass to be about 10585 MeV/$c^2$,
about 23 MeV/$c^2$ above that of Tornqvist \cite{deusons,Tornqvist:2004qy}.

\section{Strategies for observation}

The likely mixing of $X_b$ with the $\chi_{b1}(3P)$ bottomonium state suggests
that decays of the latter (and of lighter $\chi_b$ states) will provide a good
guide to isospin-conserving $X_b$ decays.  We focus on several final states.

\subsection{$X_b \to \Upsilon(1S) \omega = \Upsilon(1S) \pi^+ \pi^- \pi^0$}

This process has features in common with the decays $\chi_{b1,2}(2P) \to
\Upsilon(1S) \omega$ observed by the CLEO Collaboration \cite{Severini:2003qw}:
\bea
{\cal B}(\chi_{b1}(2P)\to\Upsilon(1S)\omega) & = &
 (1.63^{+0.35+0.16}_{-0.31-0.15})\%~, \nonumber \\
{\cal B}(\chi_{b2}(2P)\to\Upsilon(1S)\omega) & = &
(1.10^{+0.32+0.11}_{-0.28-0.10})\%.
\eea
An estimate of the rate for $\chi_{b1}(3P) \to\Upsilon(1S)\omega$ is difficult
because the increased $Q$-value may be offset by a smaller transition matrix
element.  The total width predicted for $\chi_{b1}(2P)$ \cite{Kwong:1988ae} is
79 keV, so the branching fraction quoted above corresponds to an expected
partial decay width $\Gamma(\chi_{b1}(2P)\to\Upsilon(1S)\omega) \simeq 1.3$
keV, about 1/3 of the partial decay rates for $\chi_{b1}(3P) \to \gamma
\Upsilon(1S,2S)$ predicted in Table \ref{tab:3Pbb}.  No significant signal
for $X_b \to \Upsilon(1S) \pi^+ \pi^- \pi^0$ has been seen by the Belle
Collaboration \cite{He:2014sqj}.

\subsection{$X_b \to \Upsilon(2S) \omega^* = \Upsilon(2S) \pi^+ \pi^- \pi^0$}

The $Q$-value of this decay is too small to permit the production of a real
$\omega$, but the three-pion system can still be produced with an effective
mass up to 540--560 MeV, depending on the exact mass of $X_b$.

\subsection{$X_b \to \chi_{b1} \pi^+ \pi^-$}

This decay has features in common with the decays $\chi_b(2P) \to \chi_b(1P)
\pi\pi$ observed by the CLEO \cite{Cawlfield:2005ra} and BaBar
\cite{Lees:2011bv} Collaborations.  The Particle Data Group \cite{PDG} quotes
the averages
\beq
{\cal B}(\chi_{b1}(2P)\to\chi_{b1}(1P)\pi\pi) = (9.1 \pm 1.3) \times 10^{-3}~,~
{\cal B}(\chi_{b2}(2P)\to\chi_{b2}(1P)\pi\pi) = (5.1 \pm 0.9) \times 10^{-3}~.
\eeq
Note that the total spin of the bottomonium system is preserved in these
decays, so it is reasonable to assume that will also be the case in $X_b$
decays.  There is just barely enough $Q$-value to permit the decay
$X_b \to \chi_{b1}(2P)\pi\pi$, so it makes more sense to look for
$X_b \to \chi_{b1}(1P)\pi\pi$, followed of course by $\chi_{b1}(1P) \to
\gamma \Upsilon(1S)$.

\subsection{$X_b \to \Upsilon(3S) \gamma$}

We have argued that if $X_b$ contains a substantial amount of
$\chi_{b1}(3P)$ in its wave function, this decay is likely to
dominate over $X_b \to \Upsilon(1S,2S) \gamma$.

\subsection{$X_b \to \Upsilon(1S,2S) \gamma$}

As the decays $\chi_b(3P) \to \Upsilon(1S,2S) \gamma$ have been observed
\cite{ATLASchi3P,D0chi3P,LHCbchi3Pa,LHCbchi3Pb}, and the $X_b$ is expected
to mix strongly with the $\chi_{b1}(3P)$, it is worth while to examine the
$\Upsilon(1S,2S) \gamma$ mass spectra for any departures from single
Breit-Wigner behavior.

\section{Conclusions}

We have offered several suggestions for identifying the $X_b$, the bottomonium
analogue of $X(3872)$.  We have noted the close proximity of its predicted
mass to the bottomonium state $\chi_{b1}(3P)$ recently identified at hadron
colliders.  Thus, we expect a molecular state of $B \bar B^* + {\rm c.c.}$
to mix strongly with a bottomonium state, and many of the decay modes of
$X_b$ to mirror those of a pure $\chi_{b1}(3P)$ state.  The most promising of
these include $\Upsilon(1S) \omega$, $\chi_{b1}(1P)\pi\pi$, and $\Upsilon(3S)
\gamma$.  The mass spectra in the latter final state (as well as
$\Upsilon(1S,2S)\gamma$) should show some departures from a pure Breit-Wigner
shape if examined with sufficient resolution.

{\it Note added:}  After this work was submitted for publication we became
aware of the work of Ref.\ \cite{Cardoso:2014xda}, which finds $X(3872)$ to
be a mixture of $c \bar c$ ($\sim 27\%$), $D^0 \bar D^{*0} + {\rm c.c.}$
($\sim 65\%$) and $D^+ D^{*-} + {\rm c.c.}$ ($7 \%$) with sub-percent
contributions from other states.  It would be interesting to see what a
similar approach gives for $X_b$. 

\section*{Acknowledgements}

We thank James Catmore, Simon Eidelman, Erez Etzion, Chenping Shen, and Bruce
Yabsley for discussions about the experiments, and Alexey Petrov and Qiang Zhao
for theoretical comments.  The work of J.L.R. was supported by the U.S.
Department of Energy, Division of High Energy Physics, Grant No.\
DE-FG02-13ER41958.

\end{document}